\documentstyle{EuroPhys}
\input EuroMacr
\input epsf

\begin{document}
\shorttitle{Mu{\~{n}}oz, Marini and Cafiero, Phase separation
in systems with absorbing states}
\title{ \large \bf Phase separation in systems with absorbing states.}
\author{Miguel A. Mu{\~{n}}oz $^{1}$, U. Marini Bettolo Marconi
  $^{1,2}$ and
R. Cafiero $^{1,3}$}

\institute{$^{1}$
I.N.F.M. Sezione di Roma 1, and Dipartimento di Fisica,
Universit\'a di Roma ``La Sapienza'',~Piazzale
A. Moro 2, I-00185 Roma, Italy\\
$^2$ I.N.F.M. Sezione di Camerino and 
Dipartimento di Matematica e Fisica, Universit\`a di
Camerino, Via Madonna delle Carceri, I-62032, Camerino, Italy\\
$^3$  Max-Planck Institut f\"ur Physik komplexer Systeme,
N\"othnitzer Strasse 38, D-01187 Dresden, Germany
 }
\rec{ }{ }
\pacs{
\Pacs{64}{60.-i}{General studies of phase transitions}
\Pacs{64}{60.Cn}{Order disorder transformations}
}
\maketitle

\begin{abstract}
  We study the problem of phase separation in
systems with a positive definite order parameter, and
in particular, in systems with absorbing states. Owing
to the presence of a single minimum in the free energy
driving the relaxation kinetics, there are some basic
properties differing from standard phase separation.
We study analytically and numerically this class of systems;
in particular we determine the phase diagram, the
growth laws in one and two dimensions and the presence of scale
invariance. Some applications are also discussed.
\end{abstract}
\vspace{8pt}

    The presence of conservation laws plays a key role in determining
the phenomenology and critical behavior of phase transitions
 both in equilibrium and
nonequilibrium systems. For instance, the dynamics of models
for magnetism change dramatically depending on whether the order
parameter is a conserved quantity
 or not \cite{HH}. In the conserving case, models may exhibit
 phenomena such as
spinodal decomposition (SD hereafter) and
droplet growth via condensation-evaporation
\cite{Gunton,Langer,Bray}, which are absent in models with non-conserving
dynamics.
 The problem of SD that appears, for instance in binary alloys, polymers and
spin systems with conserved magnetization,
have
attracted a lot of interest in the
last decades \cite{Gunton,Langer,Bray},
 and despite the efforts devoted to understand
their phenomenology, some of their aspects remain unclear.
 The key issue
is that of
understanding the dynamics of a system quenched from a state in the
homogeneous phase into a broken-symmetry phase. Depending on the
value of the conserved order parameter, $M$, different types of
morphologies can show up.
    The archetypal field theory representing this class of systems is
the model B (or Cahn-Hilliard equation), defined by
the following Langevin equation
 \begin{equation}
\partial_t \phi(x,t)
 =  \nabla^2 {\delta F[\phi(x,t)] \over \delta \phi(x,t)
} + \nabla \cdot {\bf \eta }(x,t)
\label{Lan}
\end{equation}
where $F[\phi(x,t)]$ is a free energy functional given by
\begin{equation}
F[\phi]= \int dx \left[ -{a \over 2} \phi^2 (x) + {b \over 4} \phi^4(x) +
 {1 \over 2} (\nabla \phi (x))^2 \right],
\label{ch}
\end{equation}
with $a$ and $b$ positive constants, and
$ {\bf \eta}(x,t)$ is a d-dimensional noise vector whose components
obey
\begin{equation}
\langle \eta_i(x,t) \rangle
=0, ~~ \langle \eta_i(x,t) \eta_j(x',t') \rangle =  D \delta^{(d)} (x-x')
\delta(t-t') \delta_{i,j}.
\label{noise1}
\end{equation}
A brief discussion of its phenomenology 
is as follows:
  an homogeneous initial condition with averaged field density
$\phi_0$, located in the 
  unstable zone of the
 free energy, experiences a linear instability
 which favors the growing of  Fourier modes with $k \neq0$ (waves),
 and the initially homogeneous phase roughens, in such a way that
locally the system tries to relax to one of the two minima of the
free energy, conserving the total order parameter. 
Let us point out that the presence of two
 different minima in Eq. (\ref{ch})
 is a key 
ingredient for phase separation.
  After an initial stage
the wave amplitude grows, until it saturates, i.e. the maxima
and minima reach the maximum and minimum of the free
energy potential, and the symmetry with respect to  the initial 
value of the field
is lost \cite{Joly}.
 Afterwards the clusters of the minority phase coalesce
via diffusive annihilation
of interfaces.
   For metastable initial conditions the
system is locally  stable, but fluctuations can take the local
order parameter to one of the minima, in this way droplets are formed,
and they grow via nucleation.

 In a seminal work Lifshitz and Slyozov
\cite{LS}, and
independently  Wagner \cite{Wagner}, derived some exact results for
very diluted quenches, i.e. when the volume fraction of one phase is
 much larger than the other one.
For example, the averaged gyration radius of clusters in the
minority phase, $R$, grows with time as $ R \approx t^{1/3}$, while
it grows like $t^{1/2}$ in the non-conserved case.
These results have been shown to be independent of the spatial dimension
$d$ \cite{indep} and confirmed in numerical studies \cite{Rogers,Raul}.
 Although the Lifshitz-Slyozov-Wagner theory (LSW)
is only valid for very dilute systems, it has been argued that a similar
law $ t^{1/3}$ is valid also for generic volume fractions
\cite{gener}. An intermediate regime has also been identified in some
cases \cite{Puri}; it is characterized by the presence of soft
domain walls and a growth law $t^{1/4}$; after this regime the
domain walls harden (become sharper) and the real asymptotic
behavior $ t^{1/3}$ sets in. In other cases in which
bulk diffusion is stopped the assymptotic domain growth law is characterized 
by an exponent $1/4$ \cite{Lacasta}.

Another important result is the identification of
a single relevant length
scale in the late stages of phase separation. It is given by the averaged
droplet size. This makes it possible to describe the late stages
of growth in terms of scaling functions
\cite{Gunton,Raul,Marro,Furukawa},
 and makes the problem suitable to be
studied using renormalization group techniques
 \cite{Mazenko,Bray}.  It is accepted that
in the scaling regime stochastic effects are irrelevant
and therefore the same behavior is obtained in purely
deterministic systems with the same symmetries and conservation
laws.

   In this paper we face the generalization of the previous
ideas to a vast class of nonequilibrium systems, namely, the
class of {\it systems with absorbing states}.
It is well established
that all the systems exhibiting a
continuous  transition
into a {\it unique} absorbing state,  without any other extra
symmetry or conservation law,
belong to the same universality class,
namely, that of directed percolation \cite{Ma,Granada}.
Among other models in that broad class are the following:
directed percolation, the contact process,
catalytic reactions on surfaces, branching annihilating
random walks,
 damage spreading and self organized criticality
 (see \cite{Ma,Granada} and references therein).
 The  Reggeon field theory (RFT) is the minimal continuous theory
in  this vast universality class \cite{Granada}.
 In its conserved version
 it is defined by a Langevin equation
analogous to Eq. (\ref{Lan})
where now $F[\phi(x,t)]$ is a different free energy functional given by
\begin{equation}
F[\phi]= \int dx \left[ -{a \over 2} \phi^2 (x) + {b \over 3} \phi^3(x) +
  {1 \over 2} (\nabla \phi (x))^2 \right],
\label{RFT}
\end{equation}
and
\begin{equation}
\langle \eta_i(x,t) \rangle
=0, ~~ \langle \eta_i(x,t)
 \eta_j(x',t') \rangle =  D \phi(x,t) \delta^{(d)}(x-x')
\delta(t-t') \delta_{i,j}.
\label{noise2}
\end{equation}
The main properties of this equation in contrast
to those of model B are:
(i) all the dynamics ceases for
$\phi(x)=0$, as corresponds to a system with an absorbing state;
(ii)
the RFT describes the dynamics of a density like,
positive definite,
order parameter, and
(iii) no up-down symmetry is present in
this type of systems and therefore the first nonlinearity in the
free energy is a
cubic term instead of a quartic one as in model B (observe, however,
that none of the forthcoming results depend on the degree 
of the non-linearity).

  Owing to these last two differences it is clear that the
essential mechanisms present in standard phase separation cannot
straightforwardly be translated to the conserved RFT: The deterministic
free energy $F$ exhibits only one minimum (at $0$ if $a \leq 0$, and at
$\phi=a/b$ if $a>0$), and the process of phase separation, if present,
does not correspond to a local relaxation of the density to the
two minima of the local free energy as in model B.

  Before proceeding further, let us present a simple toy model 
in economics that shares the properties of our previously described system
(as described by Eq. (\ref{Lan}), (\ref{RFT}) and (\ref{noise2}).
This toy model can be view as a discretization of our proposed equations,
and maybe be useful to develop an intuition of the forthcoming conclusions
and results.  
 The
model is defined on a 2-dimensional discrete lattice, 
each site representing
an individual. 
A field variable describes the wealth of each individual at each time. 
The 
wealth is updated at each time step following three different
mechanisms (1) the wealth is driven by an external agent (a fair
government) that tries to keep all the individual equally rich and 
corresponds to the potential in our model;
(2) the wealth is redistributed periodically among nearest 
neighbors ( this corresponds to the Laplacian in (\ref{RFT}) and 
describes fair citizens), and (3) in addition there are
random fluctuations. The dynamics is local (wealth flows only through nearest
 neighbors) and conserves the total wealth.
The problem is whether the individuals will remain equally rich
on average or not with these utopian rules.  As we will show 
in what follows the answer depends on the total wealth of the society.

 Coming back to the general problem,
 a straightforward analysis of local stability of our continuous model
 shows that
 the homogeneous state is
unstable with respect to fluctuations whenever $\phi < a/2b$.
  The problem arises of what is the kinetics in this case; if the
system segregates in different phases what do they correspond to?

By Fourier transforming the Langevin equation, it is easy to verify
that the fastest growing mode is
$k_m= \sqrt{ {(a- 2 b \phi_0)}/ 2 },$
and therefore the fastest growing length,
 given by $\lambda= 2 \pi /k_m$
is
$\lambda=  2 \pi / \sqrt{ {(a -2 b \phi_0)/2} }.$
The system size has to  be larger
than $\lambda$ for the instability to be observed; for finite systems
of size $L$,
the effective instability point is shifted from $ \phi_0=a/2b$
 to a smaller
value $\phi_0 \approx (a - { 2 (2 \pi)^2 \over L })/2b$, determined by the
condition $k_m =L$.

 For parameter values in the locally unstable regime 
initial homogeneous
conditions break down, and a wavy solution dominated by the
 wave vector $k_m$
 sets in.
This wavy solution cannot grow in amplitude indefinitely
due to the physical
 constraint at $\phi=0$; and, in contrast to what happens
in standard SD the maxima and minima of the
wavy solution do not collapse to the two minima of the deterministic
potential
(which in our case is monostable).
 Therefore, if
the system is unstable the question arises of what is the true stable
solution. In order to answer the same question for the model B, one can
solve the equation $\delta F[\phi] / \delta \phi =0 $ with
$F[\phi]$ given by Eq. (\ref{ch}). A well known solution of the
previous equation is $\phi(x)=\sqrt{a/b} \tanh (\sqrt a x/(2b))$. This
is a  {\it topological solution} connecting  the two minima of $F$.
In our case such a topological solution cannot exist due to the presence 
of only one minimum in $F$.
This, however, does not mean that the full stationary equation, i.e.
$\nabla^2 (\delta F[\phi] / \delta \phi )=0$ does not admit
 a topological solution, even though
in this case it is a much less intuitive one. Let us look for such a
solution, and let us do it first
 in $d=1$ (the generalization to higher dimensions being
straightforward). The general solution of
the Laplace equation, $\nabla^2 f(x)=0$ in $d=1$ is $f(x)=h+gx $
 where $h$ and $g$ are undetermined constants; therefore
a solution of the stationary problem is also a solution of
 $-{\delta F \over \delta \phi}= h + g x$ (where we have just
 replaced $f$ by $-{\delta F \over \delta \phi}$).
Imposing periodic boundary conditions, we can fix $g=0$, 
 while $h$ is specified by the
initial condition,
 $h V=-\int dx {\delta F \over \delta \phi}|_{\phi=\phi_0}$
where $V$ is the system volume (which transforms
into $h=a\phi_0-b\phi^2_0$ with $\phi_0=\phi(t=0)$
for homogeneous initial
conditions).
 Observe that the previous  equation, $-{\delta F \over \delta \phi}= h$,
 can be rewritten
as ${\delta G \over \delta \phi}= 0$, with $G[\phi]=F[\phi] + h \phi$,
and therefore the effect of the initial condition is to originate a
constant
effective field, $h$,  coupled linearly
to the field variable.  We can now look
for solutions of this modified equation, which may have two minima.
In fact, the morphology
of the new deterministic potential (in terms of $G$) is as follows:

 (i) For $\phi_0  > a/b$, h is negative, and the effective potential
has only a minimum (there is another, nonphysical extremum of
the effective potential located at some negative value of $\phi$).

 (ii) At $a/b \ge \phi_0 \ge a/2b$
 the effective potential  develops a maximum at $\phi_M= a/b-\phi_0$,
and the minimum is at $\phi_m=\phi_0$.
 Between $\phi=0$
and $\phi_M$ the potential is a monotonously increasing function of $\phi$,
therefore at $\phi=0$ the potential exhibits a relative minimum
even though its derivative is not zero. As the value of $\phi_0$
decreases the maximum and the minimum approach each other.

 (iii) For $a/2b \ge \phi_0 \ge 0$, the two extrema (that collapse
into a saddle point at $\phi_0= a/2b$) start separating again as 
$\phi_0$ is reduced.
The maximum, located now at $\phi_M=\phi_0$
 returns to $0$ at $\phi_0=0$, while the minimum,
$\phi_m= a/b-\phi_0$, increases as $\phi_0$ is decreased.

  Observe that in the interval $[0, a/b]$ the effective potential
has the same form at $a/2b + \Delta$ and at $a/2b-\Delta$. 
Note also that in that interval the potential has two minima,
one with zero derivative, and another at the origin with, in general,
 non-vanishing
derivative, which is due to the presence of an absorbing
barrier at $\phi=0$.
  It is also possible to decide the fraction of total density
located at each minima; supposing in first approximation
 that the
potential is locally minimized at all the points simultaneously, 
and
calling $p$ the fraction at $\phi = 0$ and $1-p$ the complementary
probability in such a way that $\phi_0 = 0 p  + (1-p) \phi_m$, we obtain
$p=(a-2 b \phi_0) / (a- b \phi_0)$. In particular the  {\it mixture}
is symmetric at $\phi_0=a/(3b)$.

 We have performed some numerical simulations to further study the proposed model.
For that we consider an initial condition with average field density $\phi_0$ 
and small fluctuations around that value. 
 In figure 1 we show a long time typical density profile in $d=1$
 ($L=100$) for two different initial densities in the unstable regime:
$\phi_0 =
0.20$ and $\phi_0 =
0.26$.
The two phases are 
clearly asymmetric; the phase at $\phi=0$ is free of fluctuations.
At $\phi_0=0.26$ ( i.e. slightly below the critical point of instability,
$a/2b=0.3$ for the chosen parameter values)
 the fraction of sites in the nonvanishing region is larger
than in the other case, but in any case, the walls are quite soft.

\begin{figure}[h]
\epsfxsize=5cm
\epsffile{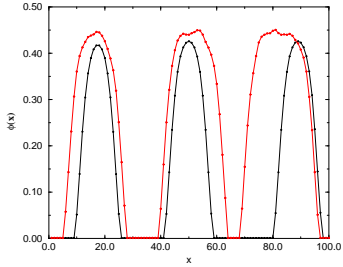}
\caption{Density profile in the one-dimensional case for $\phi_0 =
0.20$ (lower curve) and $\phi_0 =
0.26$ (upper curve) respectively at $t \approx 10^6$.
 The parameter values are taken
to be: $a=0.6$, $b=1$, and $D=0.01$.
}
\label{fig1}
\end{figure}

We have verified the presence of only one characteristic
length is the scaling regime. For that we have computed the
structure function at different times, and rescaled it using
 $ R(t)$; a good collapse of the curves is
observed (see Fig.2).

\begin{figure}[h]
\epsfxsize=6cm
\epsffile{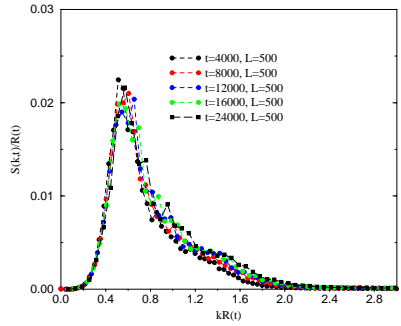}
\caption{Rescaled structure function in $d=1$ for times
 from $t=4000$ to $t=24000$.
}
\label{fig3}
\end{figure}

  For the time evolution of the averaged cluster size we
get, in $d=1$: $R(t) \sim t^{(\alpha \pm 0.02)}$, 
 with $\alpha= 0.30, 0.28$ and $0.26$, for initial
concentrations $\phi_0= 0.05, 0.1$ and $0.15$ respectively (Fig.3).

\begin{figure}[h]
\epsfxsize=6cm
\epsffile{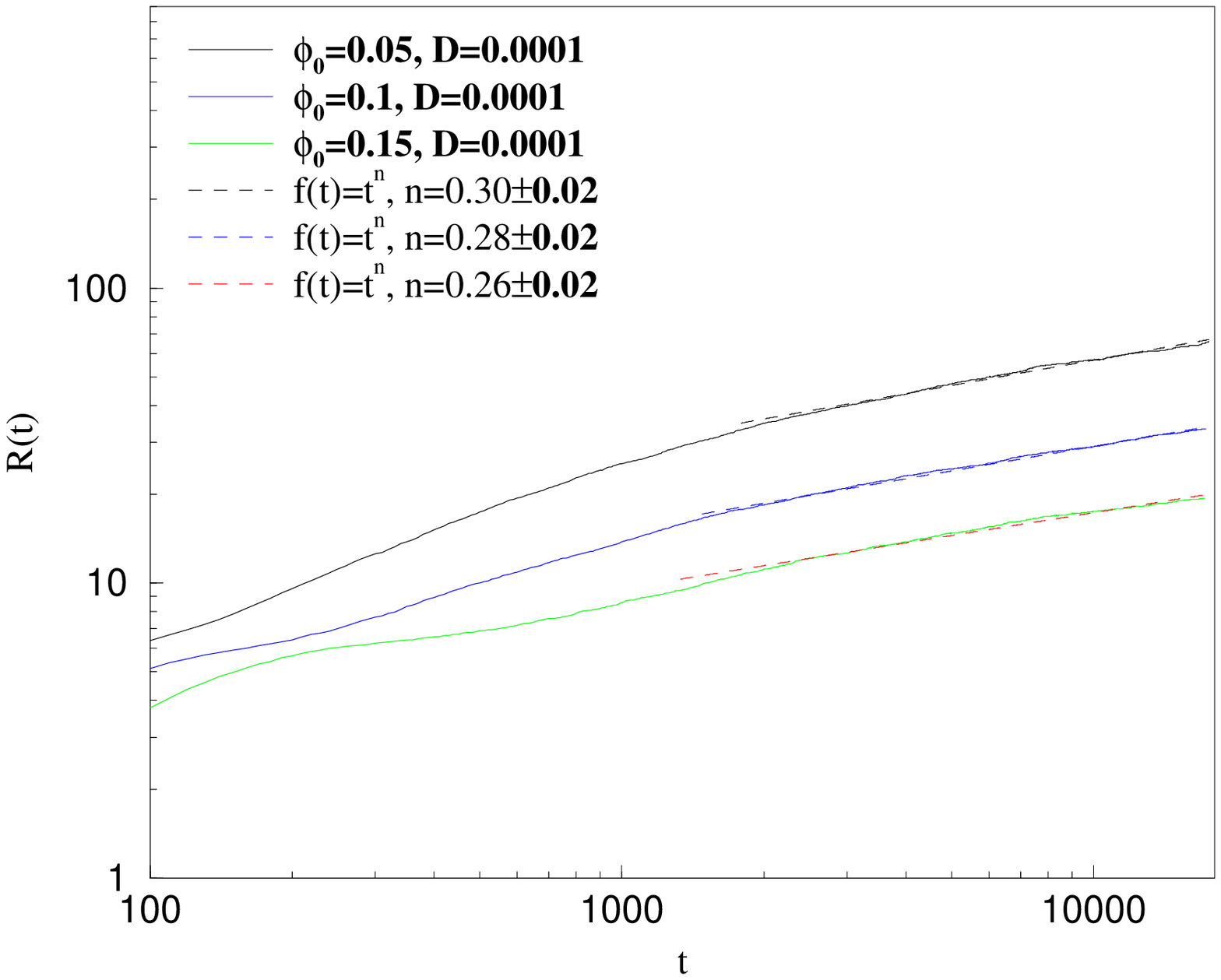}
\caption{Log-log plot of the averaged mean cluster radius
as a function of time for $d=1$, and different
initial densities.
}
\label{fig2}
\end{figure}

  Deciding whether the assymptotic behavior is governed by an exponent
 $1/3$ or $1/4$ is
a difficult task from both theory and numerics.
From the theoretical point 
of view it has been argued that when the bulk dynamics is suppressed
an exponent $1/4$ 
can be obtained \cite{Lacasta}, while time dependent exponents,
interpolating between $1/3$ and $1/4$ (reflecting crossover effects)
are obtained for concentration dependent diffusion coefficient \cite{Lacasta}.
Observe that in our 
model the dynamics ceases in the interior of absorbing clusters
but not in the other (active) phase. 
 More systematic
simulations and further theoretical analysis aiming to sort out this point
 will be presented elsewhere.

 In $d=2$ a similar qualitative behavior is obtained. In Fig.4 we
show the evolution of the field variable as a function of time for
an initial density $\phi_0=0.05$. Observe that first absorbing (black) 
regions are created in a sea of active sites (white) (at time $t=0$ the
whole lattice would be white). Afterwards these islands 
percolate through the system and the active regions become isolated.

\begin{figure}[h]
\epsfxsize=7cm
\epsffile{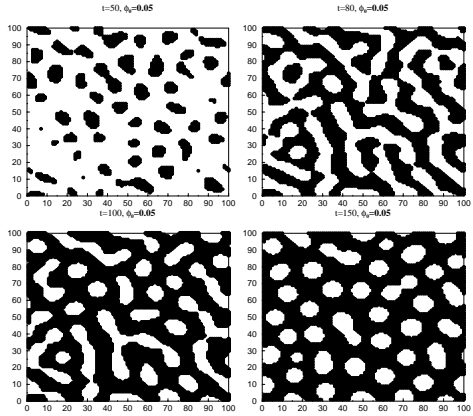}
\caption{Time evolution of the density in a $100 \times 
100$ lattice  for an initial 
average concentration 
$\phi_0=0.05$ (with small fluctuations around that mean value).
Black points represent sites in the absorbing state, while active regions 
(i.e. above a certain tiny threshold) are represented in black. First
the initial condition in unstable and absorbing regions are created ($t=50$). 
Then these regions coalesce ($t=80$) and percolate through the system, isolating
regions of active sites ($t=100$). The size of active regions
 shrinks down for late times
($t=150$). }
\label{fig4}
\end{figure}

  All the measured effects in both $d=1$ and $d=2$,
 are essentially deterministic, 
and correspond therefore to the $T=0$ fixed point of the 
conserved RFT. Therefore, the main ingredient characterizing
this phase separation in systems with an absorbing state,
is the physical constraint at $\phi=0$, and not the noise 
itself.
 
  Our system may exhibit a phase
transition from the ordered behavior to a disordered phase 
as the noise amplitude is increased. 
 Due to the presence of a noise of a different nature
we expect this transition
to be in a universality class other than that of model B.
 This
issue will be analyzed in a forthcoming paper. 

  Summing up, we have shown that 
deterministically monostable systems, and, in particular, systems 
with absorbing barriers can 
exhibit the phenomenon of phase separation, in a form
that shows many common features with the standard picture 
of phase separation, but has also some essential differences.
These come mainly from the fact that the two minima of the effective 
potential are intrinsically of different nature in our 
model.

 Among other many possible applications of the general 
phenomena 
we have just described let us cite the following: phase
separations of charged particles in a conductor 
(observe that these situation has an absorbing phase: zones with
no charge are fluctuation free), or diffusion of particles 
in an absorbing substrate. We expect the general
scenario studied here to describe some realistic model 
similar to the ones cited previously.

 Finally,  coming back to our toy model in economics,
the wealth remains homogeneously distributed among
 individuals when the total wealth in the society 
is larger than a
critical value,
(determined by the government tax policy),
 but when it is smaller than the critical
threshold the society becomes separated in two 
groups: rich and poor, which separate from each other.

It is a pleasure to acknowledge useful discussions with J. M. Sancho
J. Marro, P. Garrido, L. Pietronero, and R. Toral.
This work has been partially
supported by the European Union through a grant to M.A.M.
(contract ERBFMBICT960925).


\end{document}